\documentclass{aa}  

\usepackage{natbib}
\bibpunct{(}{)}{;}{a}{}{,} % to follow the A&A style

\usepackage{xcolor}
\usepackage{subfigure}

\usepackage{graphicx}
\usepackage{txfonts}
\usepackage[normalem]{ulem}

\begin{document}

   \title{\textsc{Forklens}\thanks{https://github.com/zhangzzk/forklens}: Accurate weak-lensing shear measurement with deep learning}

   \author{Zekang Zhang\inst{1,2}, Huanyuan Shan\inst{1,2,3}\thanks{\email{hyshan@shao.ac.cn}}, Nan Li\inst{4,11}\thanks{\email{nan.li@nao.cas.cn}}, Chengliang Wei\inst{5}, Ji Yao\inst{1}, Zhang Ban\inst{6}, Yuedong Fang\inst{7}, Qi Guo\inst{4}, Dezi Liu\inst{8}, Guoliang Li\inst{5}, Lin Lin\inst{1}, Ming Li\inst{4}, Ran Li\inst{4,10,11}, Xiaobo Li\inst{6}, Yu Luo\inst{5}, Xianmin Meng\inst{4}, Jundan Nie\inst{4}, Zhaoxiang Qi\inst{1,11}, Yisheng Qiu\inst{9}, Li Shao\inst{4}, Hao Tian\inst{4}, Lei Wang\inst{5}, Wei Wang\inst{6}, Jingtian Xian\inst{6}, Youhua Xu\inst{4}, Tianmeng Zhang\inst{4}, Xin Zhang\inst{4}, Zhimin Zhou\inst{4}}

   \institute{Shanghai Astronomical Observatory, Chinese Academy of Sciences, Shanghai 200030, PR China
        \and
        University of Chinese Academy of Sciences, Beijing 100049, PR China
        \and
         Key Laboratory of Radio Astronomy and Technology, Chinese Academy of Sciences, A20 Datun Road, Chaoyang District, Beijing, 100101, P. R. China
        \and 
        National Astronomical Observatories, Chinese Academy of Sciences, Beijing 100101, PR China
        \and 
        Purple Mountain Observatory, Chinese Academy of Sciences, Nanjing, 210023, PR China
        \and
        Changchun Institute of Optics, Fine Mechanics and Physics, Chinese Academy of Sciences, Changchun, 130033, China
        \and
        University Observatory, Faculty of Physics, Ludwig-Maximilians-Universität, Scheinerstr. 1, 81679 Munich, Germany
        \and
        South-Western Institute for Astronomy Research, Yunnan University, Kunming, 650500, China
        \and
        Research Center for Astronomical Computing, Zhejiang Laboratory, Hangzhou 311100, China
        \and
        Institute for Frontiers in Astronomy and Astrophysics, Beijing Normal University,  Beijing 102206, China
        \and
        School of Astronomy and Space Science, University of Chinese Academy of Science, Beijing 100049, China
             }

%   \date{Received September 15, 1996; accepted March 16, 1997}

% \abstract{}{}{}{}{} 
% 5 {} token are mandatory
 
  \abstract
  %{} leave it empty if necessary  
  {Weak gravitational lensing is one of the most important probes of the nature of dark matter and dark energy. In order to extract cosmological information from next-generation weak lensing surveys (e.g., \textit{Euclid}, \textit{Roman}, LSST, and CSST) as much as possible, accurate measurements of weak lensing shear are required.}
  {There are existing algorithms to measure the weak lensing shear on imaging data, which have been successfully applied in previous surveys. In the meantime, machine learning (ML) has been widely recognized in various astrophysics applications in modeling and observations. In this work, we present a fully deep-learning-based approach to measuring weak lensing shear accurately.}
  {Our approach comprises two modules. The first one contains a convolutional neural network (CNN) with two branches for taking galaxy images and point spread function (PSF) simultaneously, and the output of this module includes the galaxy's magnitude, size, and shape. The second module includes a multiple-layer neural network (NN) to calibrate weak-lensing shear measurements. We name the program \textsc{Forklens} and make it publicly available online.}
  {Applying \textsc{Forklens} to CSST-like mock images, we achieve consistent accuracy with traditional approaches (such as moment-based measurement and forward model fitting) on the sources with high signal-to-noise ratios (S/N, > 20). For the sources with S/N < 10, \textsc{Forklens} exhibits an $\sim 36\%$ higher Pearson coefficient on galaxy ellipticity measurements.}
  {After adopting galaxy weighting, the shear measurements with \textsc{Forklens} deliver accuracy levels to 0.2\%. The whole procedure of \textsc{Forklens} is automated and costs about $0.7$ milliseconds per galaxy, which is appropriate for adequately taking advantage of the sky coverage and depth of the upcoming weak lensing surveys.}

   \keywords{cosmology:observations --
            gravitational lensing: weak --
            methods: data analysis
            }

    \titlerunning{\textsc{Forklens}}
    \authorrunning{Zhang, Shan et al.}
    % \nocopyright
    \maketitle

   % \titlerunning
%
%-------------------------------------------------------------------
% \footnotetext{https://github.com/zhangzzk/forklens}
\section{Introduction}

Gravitational lensing is a phenomenon that describes the deflection of light from background sources by the gravitational potential of matter. It has become one of the most promising tools for the study of various topics in astrophysics, as it is directly sensitive to the distribution of matter, including both dark and visible matter. In the weak -lensing regime, the deflection distortions account for only a few percent of the object's intrinsic shape, which is also called shear. By measuring the spatially correlated shears of an ensemble of galaxies, one can map the mass profile of galaxy clusters, identify voids, and even probe the large-scale matter distribution of the Universe. By further considering galaxy redshifts, weak lensing is also used to study the growth of structure and the nature of dark energy (for a recent review on weak gravitational lensing, see \citealt{mandebaum2018review}).

Since the first detection made decades ago \citep{bacon2000wl,kaiser2000wl}, cosmic shear has matured into an important approach for cosmological surveys. Several large surveys have now been put into action, including the Kilo Degree Survey (KiDS, \citealt{hildebrandt2017kids}), the Dark Energy Survey (DES, \citealt{krause2017des}), and the Subaru Hyper SuprimeCam lensing survey (HSC, \citealt{aihara2018hsc}). There are also several upcoming experiments, such as {\it Euclid} \citep{laureijs2011euclid}, the Nancy Grace Roman Space Telescope (Roman, \citealt{spergel2015roman}), the Vera C. Rubin Observatory (LSST, \citealt{abell2009lsst}), and the Chinese Space Station Telescope (CSST, \citealt{zhan2011csst, zhan2021csst}).

The methods of weak-lensing shear measurement are usually tested in simulations mimicking the real observations, where the galaxy images are sheared by a known value $\textit{\textbf{g}}_{\rm true}$. The bias resulting from various systematics  between the estimated shear $\textit{\textbf{g}}$ and the true signal is conventionally described approximately as a linear model with an additive bias $\textit{\textbf{c}}$ and a multiplicative bias $\textit{\textbf{M}}$ \citep{heymans2006linear,massey2007linear},
\begin{equation}
    \hat{\textit{\textbf{g}}}-\textit{\textbf{g}}_{\rm true} = \textit{\textbf{Mg}}_{\rm true}+\textit{\textbf{c}}
,\end{equation}
where $\textit{\textbf{g}}$ is a two-component quantity, and so is $\textit{\textbf{c}}$. $\textit{\textbf{M}}$ is a $2\times2$ matrix, while typically the off-diagonal elements are negligible and the diagonal elements are approximately the same. In this work, we examine our methods on an average of multiplicative bias $m$ and additive bias $c$. The requirement for Stage IV weak-lensing experiments (e.g., {\it Euclid}) gives that $|m|\lesssim2\times10^{-3}$ and $|c|\lesssim2\times10^{-4}$ \citep{massey2013require}.

One of the major sources of systematics is the effect of point spread function (PSF) from either the atmosphere or the optical effect of the telescope itself, while pixel response and charge diffusion may also be taken into consideration. PSF smears the shape of observed galaxies and dilutes the shear estimate which causes a multiplicative bias. PSF anisotropy also affects the measured galaxy ellipticity (i.e., PSF leakage), causing an additive bias. This requires careful treatment and correction when inferring precise weak lensing distortion \citep{paulin2008psf,paulin2009psf}.

For shear measurement, there have traditionally been two approaches: (1) one measures the weighted quadruple moments of the image light profile \citep{kaiser1995KSB,Rhodes2000moments,melchior2011moments}; (2) one fits the image assuming a galaxy and PSF model \citep{massey2005model,nakajima2007model,miller2013model}. Both moments-based methods and model-fitting methods can produce quite accurate estimates of galaxies with a high signal-to-noise ratio (S/N), but they suffer significant "noise bias", which is difficult to predict. The pixel noise can translate into a complicated and skewed distribution of the measured ellipticity \citep{melchior2012noise}, which then propagates into bias in shear estimation. It can be seen as a function of image S/N, galaxy size, galaxy shape, and surface brightness, and also the PSF shape if not well corrected. These methods have been well applied in previous surveys where a magnitude threshold on galaxy samples is introduced. This could lead to further selection bias and lower the galaxy number density when inferring shear correlations. Some of the previous works derived the function of these properties in simulation and apply the calibration to survey data \citep{miller2013model,kuijken2015calibration,jarvis2016leakage,hoekstra2015calibration}. KiDS \citep{fenech2017selection, hildebrandt2017kidsconstraint} used the "self-calibration" technique, which is directly operated on the measurements instead of simulation. Another widely recognized new method is \textsc{metacalibration} \citep{huff2017METACALIBRATION,sheldon2017}, which is based on an early similar idea by \cite{kaiser2000wl}. \textsc{metacalibration} introduces a tiny artificial shear directly on the observed image and calculates the shear response of the observed galaxy ellipticity. This method has been validated on various simulations and shows good accuracy to a cut at $\rm S/N\sim5,$ with a specific formalism in place to deal with selection bias \citep{sheldon2017}. An updated version named \textsc{metadetection} \citep{sheldon2020metadetect} is further proposed to account for the effect of multisource blending, although it does not provide a solution for redshift de-blending \citep{maccrann2022blending}. There are also some methods able to reach sub-percent level accuracy without calibration using external simulations, for example, Bayesian Fourier Domain (\textsc{BFD}, \citealt{bernstein2016BFD}), \textsc{Fourier\_Quad} (\citealt{zhang2019ft,li2021Fourier_Quad}), and Fourier power function shapelets (\textsc{FPFS}, \citealt{li2022FPFS,li2022response}).

With the fast-increasing resolution and field of view of next-generation surveys, considerably larger ensembles of galaxies across the sky are available for precision cosmology. Machine learning (ML) algorithms are specifically designed to handle large amounts of data and are optimized for efficiency, making them highly suited for tasks such as data processing and predictive modeling. Hence, over the past decade, ML has been used for a wide range of applications in gravitational lensing, from shear measurement (neural network, NN: \citealt{gruen2010ml,tewes2019ml, pujol2020calibration}; Hopfield neural network, HNN: \citealt{nurbaeva2015mlshear}; convolutional neural network, CNN: \citealt{ribli2019shape,springer2020mlshear}), convergence map and mass reconstruction (Generative neural network, GNN: \citealt{shirasaki2019mlmass}; U-Net: \citealt{jeffrey2020mlmass}), lensing modeling, and simulation (CNN: \citealt{pearson2019strong,pearson2021strong}; GNN: \citealt{lanusse2021gnn}), to cosmological constraints (CNN: \citealt{fluri2018mlconstraint,ribli2019mlconstraint,lu2022mlconstraint}).

A CNN has been used to infer shape information directly from images at the pixel level. \citet{ribli2019shape} developed a 13-layer CNN to measure galaxy shapes and apply it to the DES Y1 catalog, showing better consistency with CFHTLenS shapes. Multilayer fully-connected NNs have been used to emulate the relation between shear bias and observed galaxy properties. \cite{gruen2010ml} first proposed to let NNs analyze the data and estimate shear after training them on simulations with known shear, using the ellipticity measurement from a specific method (e.g., KSB) and further parameters that might be indicative of the bias calibration (such as galaxy size and magnitude). With similar motivation, \cite{tewes2019ml} used the same network to perform shear estimation in the presence of various feature noises such as instrument effects, unknown galaxy morphology, and image noise. They took moment-based measurements on an individual galaxy's shape as input, including ellipticity, flux, radial extension, and the concentration of the light profile. The network outputs the shear estimator based on these noisy galaxy features. Instead of minimizing the original mean squared error between target shear and galaxy features, they formulate a mean square bias (MSB), which favors the accuracy in the predictions of the explanatory variables. The training data are then carefully structured so that the NN is able to learn the general relation of the shear estimator to various galaxy realizations. Both works did not use ML on the pixelated light distribution of the galaxy, but integrated the NN-based calibration with traditional methods such as KSB. A similar idea can also be seen in \cite{pujol2020calibration}, where a multilayer NN was used to map the relation between the measured image properties of an individual galaxy and the shear bias.

In this work, to measure weak lensing shears in an automated and efficient manner for the next generations of surveys, such as Euclid and CSST, we propose the \textsc{Forklens} method, including a fork-like deep CNN that takes both galaxy and PSF images as input to measure the galaxy's shape and an artificial NN to calibrate the shear bias. The fork-like network is similar to those of \citet{maresca2021fork} and \textsc{GaLNets} \citep{li2022galnet}, which are for identifying unphysical modeling results of strong lenses and predicting the Sersic parameters of galaxies, respectively. 

% \textcolor{green}{Applying Forklens to CSST-like mock images, we achieve consistent accuracy with traditional approaches (such as moment-
% based measurement and forward model fitting) on the sources with high signal-to-noise ratios (S/N). For the sources with meagre S/N, Forklens exhibits superior predictions on galaxy shapes.}

The paper is organised as follows: we introduce our network architecture and method in Sec. \ref{method} and show the simulations for CSST as well as the training data organization in Sec. \ref{datasets}; the results of shear measurements and the comparison to traditional methods are presented in Sec. \ref{results}; finally, Sec. \ref{conclusion} lists the conclusions.

%--------------------------------------------------------------------

\section{Methodology}
\label{method}

\begin{figure*}
    \centering
    \includegraphics[width=\linewidth]{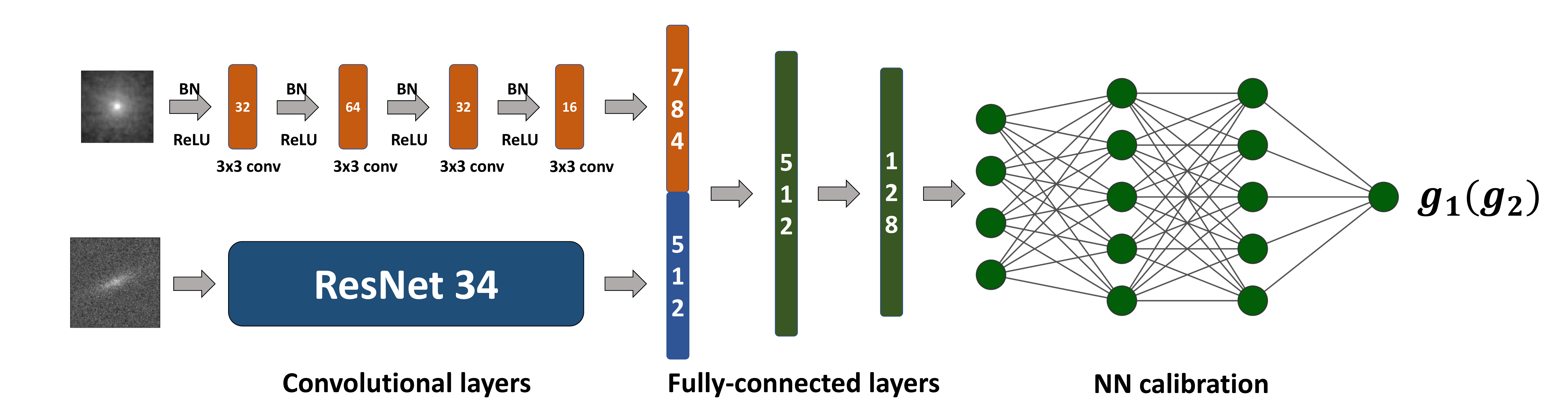}
    \caption{Outline of \textsc{Forklens} shear estimation architecture. The CNN part contains two branches, one fed with PSF and one fed with a galaxy image. The PSF branch has four convolutional layers, each with batch normalization and ReLU activation function. We adopted a 34-layer residual network to extract the information of galaxies where the image is larger and suffers from pixel noise. The two branches are then concatenated following two fully connected layers, where the effect of PSF is corrected. CNN then outputs the galaxy's properties including size, magnitude, and ellipticities. A further NN calibrates the measured features biased by noise and outputs the final shear estimate. The NN part is in practice a committee of eight independent NNs and the $g_1(g_2)$ is the average of the eight outputs.}
    \label{fig:forkcnn}
\end{figure*}

\textsc{Foklens} is a fully deep-learning-based method composed of two parts. The outline of the architecture is shown in Fig. \ref{fig:forkcnn}. In the first part, we used a CNN to measure an individual galaxy's shape (together with its size and magnitude) from the pixelated image and simultaneously corrected the effect of PSF smearing. Based on the CNN measurement, we then used an NN to estimate the shear response of the galaxy and perform calibration.

\begin{figure}
    \centering
    \includegraphics[width=\linewidth]{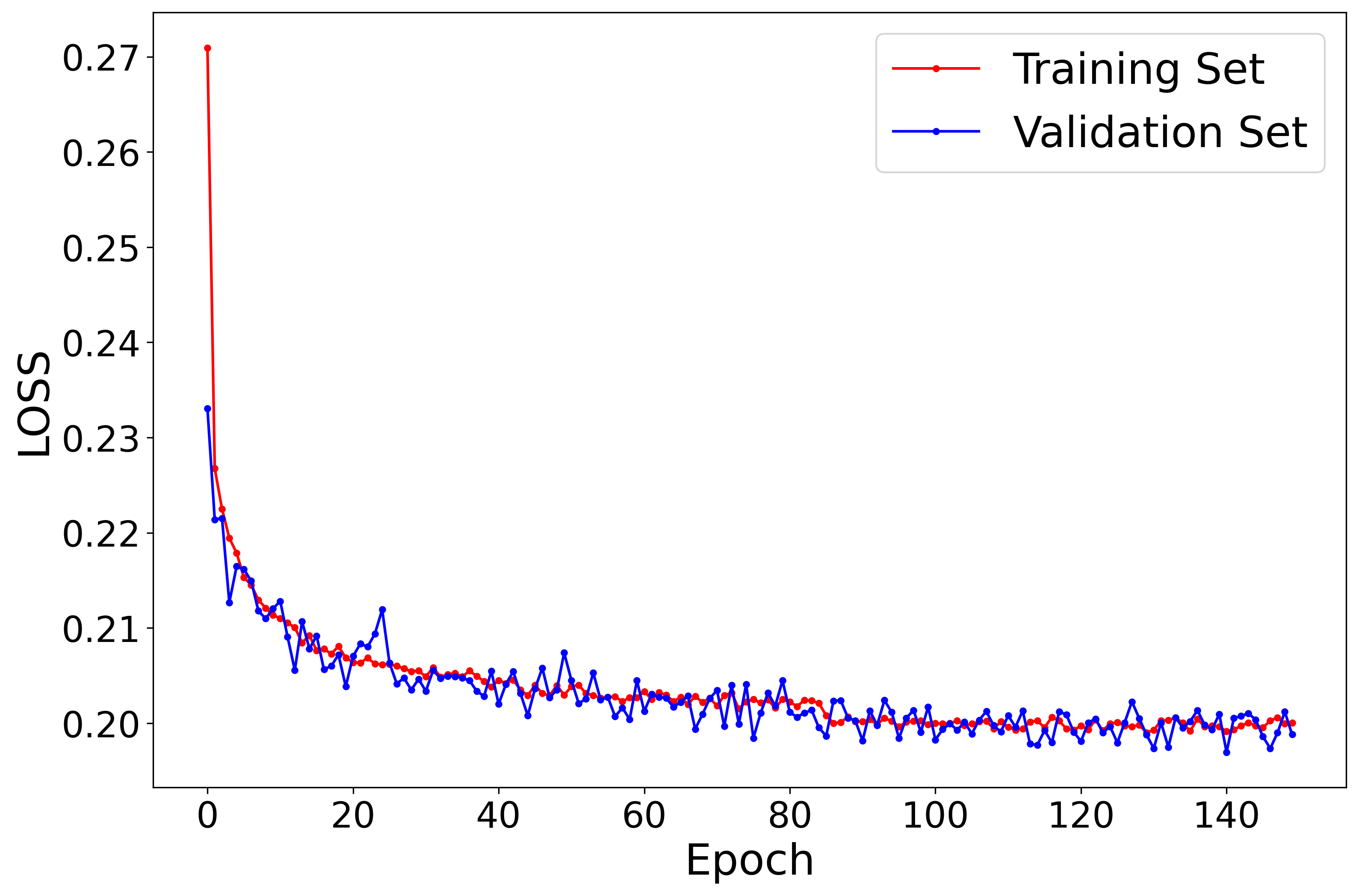}
    \caption{Training and validation LOSS of CNN. 20,000 galaxy and PSF pairs were used in total with a 10\% validation split, and the batch size is 200. The initial learning rate is 0.01 and is reduced by 0.1 times when the LOSS (Eq. \ref{mse}) has stopped improving. We take the model at the 600th epoch as our best one.}
    \label{fig:loss}
\end{figure}

\subsection{CNN architecture for shape measurement}

One of the major challenges in weak-lensing analysis is to accurately measure the shapes of small, faint galaxies, which are usually overwhelmed by observational noise. In practice, one can never perfectly measure the ellipticities as various noises can undermine the ability to extract the true shape information of the object. \cite{viola2014bias} listed three origins of measurement bias: 1) model bias when an incorrect galaxy model is adopted to describe the observation; 2) bias for the shape measurement algorithm; 3) bias introduced due to observation noise, which is caused by the nonlinearity of galaxy morphology parameters in the image pixels. All methods to measure galaxy shapes are sensitive to noise bias, even at a high S/N. Deep learning (DL) algorithms have demonstrated exceptional performance in detecting patterns in images and are capable of making more reliable predictions by mitigating the effects of noise in the input data. Although unsupervised learning has been seen in many astrophysics applications -for example, astronomical object identification \citep{han2022unsupervised,wei2022unsupervised}- supervised learning is more widely adopted in regression problems. In this case, the trained model may heavily rely on the assumed model used in the simulation and training set. Model bias therefore still requires careful treatment.

We built a custom fork-like CNN architecture with two input paths, one for the observed galaxy ($128\times128$ pixel stamp) and one for the PSF ($48\times48$ pixel stamp). We adopted ResNet34 for the first path and the second path consists of four convolutional layers, four batch normalization layers, and four layers of the rectified linear unit (ReLU) activation function. The final layers of both paths are flattened and concatenated, before being fed into three fully-connected layers. The final layer outputs a four-node array ($n_{\rm fea}=4$), which represents the predicted properties of the galaxy before PSF convolution including galaxy half-light radius and its magnitude in the i band, and two components of its ellipticity, $e_1$ and $e_2$,
\begin{equation}
    e_1+i e_2 = \frac{a-b}{a+b}\exp{i 2\theta}
,\end{equation}
where $a$ ($b$) is the length of the galaxy's semi-major (semi-minor) axis, and $\theta$ is the position angle. The four outputs are then fed into another NN for unbiased shear estimates.

Deep layers of CNN are believed to progressively learn more complex features. A growing depth has been required for accurate predictions of both image classification \citep{krizhevsky2012deep,zeiler2014deep} and regression \citep{lat2020deep} tasks. However, normal deep networks are generally hard to train and face problems such as a vanishing or exploding gradient. ResNet \citep{he2015resnet} is constructed by a series of "residual blocks" that differ from normal layers with a skip connection or a "shortcut". Such a shortcut directly adds the input of a block to its output, which makes training much deeper networks possible. ResNet34 consists of 16 residual blocks with 34 convolutional layers in total, and our results see no improvement in adopting a deeper ResNet.

To train the CNN, we input a mini-batch size of $n_{\rm bat}=200$ for 600 complete iterations on the whole training set, namely 600 epochs. We used stochastic gradient descent (SGD) as the parameter optimizer. The learning rate is initially set as 0.1 and is reduced by a magnitude when the metric has stopped improving. We adopted the mean-squared error between each element in the input labels $x$ and the predicted outputs $y$ as our LOSS function,
\begin{equation}
    {\rm LOSS} = \frac{1}{n_{\rm bat}}\sum_{i=1}^{n_{\rm bat}}\frac{1}{n_{\rm fea}}\sum_{n=1}^{n_{\rm fea}}(x_{n,i}-y_{n,i})^2
\label{mse}
,\end{equation}
which is averaged over elements every mini-batch. In this work, we did not use any hyperparameter optimization to tune the CNN architecture. Fig. \ref{fig:loss} shows the training and validation LOSS. We make \textsc{Forklens} publicly available\footnote{\url{https://github.com/zhangzzk/forklens}}.

In order to evaluate the performance of the fork-like CNN architecture, we also run other well-tested methods on the same data including moments-based measurement and model fitting to make comparisons.
For shapes based on moments, we used the EstimateShear function in the HSM module of {\it GalSim} software package\footnote{\url{https://github.com/GalSim-developers/GalSim}} \citep{rowe2015galsim}. There are several algorithms included in the function re-implemented from different works (e.g., BJ by \citealt{bernstein2002method}, LINEAR and REGAUSS by \citealt{hirata2003method}). We find their performances are quite similar, and we adopted the REGAUSS option throughout this paper.

For model fitting, we employed the route implemented in the {\it Ngmix} software package\footnote{\url{https://github.com/esheldon/ngmix}} \citep{sheldon2015ngmix}. The galaxy is fit to a single Gaussian convolved by another single Gaussian representing the PSF. {\it Ngmix} provides other more complicated models with multiple Gaussians, yet no apparent improvement is seen adopting these models and the speed is much slower with more free parameters to fit. Following \cite{zuntz2018desy1shape}, we adopted flat priors on all model parameters except the prior on ellipticity, which is the isotropic unlensed distribution as the Eq.~24 in \cite{BA2014} with $\sigma=0.1$.

\subsection{Neural network architecture for shear calibration}

% In common with other existing methods, the CNN ellipticity measurements are biased by pixel noise. As the ellipticity of an object is a non-linear quantity, even a symmetric distribution of noise will translate into a complicated and skewed distribution of the ellipticity \citep{melchior2012noise}, which propagates into shear estimation as a bias. It can be seen as a function of image S/N, galaxy size, shape, and surface brightness, and also the PSF shape if not well corrected. Some of the previous works derived the function of these properties in simulation and apply the calibration to survey data \citep{miller2013model,kuijken2015calibration,jarvis2016leakage,hoekstra2015calibration}. KiDS \citep{fenech2017selection, hildebrandt2017kidsconstraint} used the "self-calibration" technique which is directly operated on the measurements instead of simulation. Similarly, \textsc{metacalibration} introduces a synthetic shear signal into real data and derives a shear response for each image, defined as follows.

\begin{figure}
    \centering
    \includegraphics[width=0.5\linewidth]{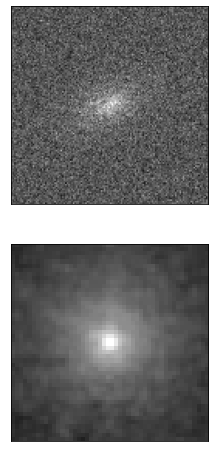}
    \caption{CSST simulation of example galaxy (top, half-light radius of 1.2 arcsec, 20 in magnitude, $e_1=0.4$ and $e_1=-0.4$) and PSF in log scale (bottom, drawn from random positions on the CCD in simulation). Both shot noise and Gaussian noise are included in the observed galaxy. PSF is simulated based on the optical design model (Sec. \ref{simulation}).}
    \label{fig:psf}
\end{figure}

\begin{figure}
    \centering
    \includegraphics[width=0.75\linewidth]{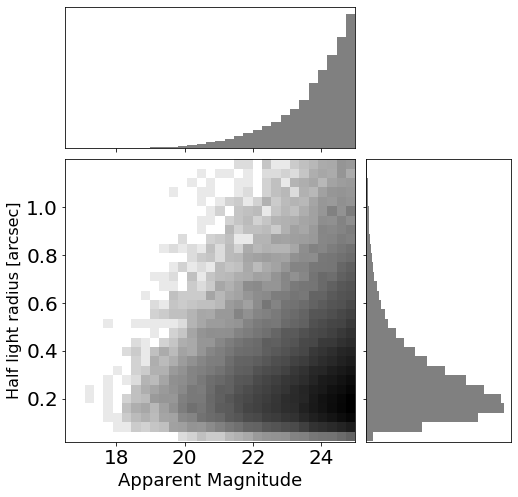}
    \includegraphics[width=0.75\linewidth]{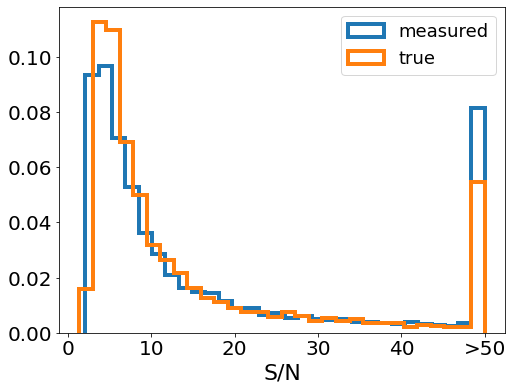}
    \caption{Distribution of simulated CSST galaxies from which we randomly drew for training and testing sets (with different random seeds). Top: Distributions of galaxy half-light radii and magnitude. Bottom: Normalized distribution of measured and true galaxy S/N (Eq. \ref{eq:snr}). Galaxies with S/N > 50 sit in bin $\sim 50$.}
    \label{fig:catalog}
\end{figure}

As with other existing methods, the CNN ellipticity measurements are biased by pixel noise. A noisy measurement of image ellipticity $\textit{\textbf{e}}$ can be expanded in a Taylor series about shear $\textit{\textbf{g}}$,
\begin{equation}
    \textit{\textbf{e}} = \textit{\textbf{e}}|_{\textit{\textbf{g}}=0}+\frac{\partial \textit{\textbf{e}}}{\partial \textit{\textbf{g}}}|_{\textit{\textbf{g}}=0} \textit{\textbf{g}} + ..., 
\end{equation}
where the first-order term is called the shear response,
\begin{equation}
    \textit{\textbf{R}}=\frac{\partial \textit{\textbf{e}}}{\partial \textit{\textbf{g}}}|_{\textit{\textbf{g}} = 0}, 
\end{equation}
and the zero-order will be statistically cancelled out over an ensemble of galaxies assuming their intrinsic shapes are randomly oriented. \textsc{metacalibration} derives the response $\textit{\textbf{R}}$ by applying an artificial shear to observed images and calculating the changes in the measurement of $\textit{\textbf{e}}$. 

In this work, we followed the same formula as \cite{tewes2019ml} to perform shear calibration, but instead took the measurement of our CNN as input. Four values are fed into the NN (including two components of ellipticities, galaxy half-light radius, and apparent magnitude), which are further passed into two hidden layers of five nodes each and output the estimator of g1 (g2). Activation functions in all input and hidden nodes are the hyperbolic tangent $f(x)=\tanh(x)$. In the output layer, it is an identity activation function. All inputs are normalized in an interval of [-1,1].

The network is optimized by minimizing the MSB loss function with a Broyden-Fletcher-Goldfarb-Shanno (BFGS) iterative optimization algorithm, 
\begin{equation}
    {\rm MSB}=\frac{1}{n_{\rm case}}\sum_{i=1}^{n_{\rm case}}[\frac{1}{n_{\rm rea}}\sum_{j=1}^{n_{\rm rea}}\hat{g}_{ij}(\textbf{\textit{f}})-g_{i}^{\rm true}]^2
\label{msb}
,\end{equation}
where $\textbf{\textit{f}}$ denotes the input galaxy features.

The network parameters are initialized randomly from a normal distribution, and different initializations will lead to different shear estimate outputs. To exploit this stochastic behaviour, a committee of eight identical but independent NNs is trained with the same data. The final shear estimate is the average over the outputs of the best four NN members (according to their training loss).

\subsection{Weighting galaxies}

In weak lensing surveys, not all of the detected sources are utilized in shear measurements or for further scientific analysis. Specific criteria are employed to select galaxies based on certain characteristics, aiming to avoid unreliable measurements and mitigate biases. One common consideration is that faint galaxies tend to have more noisy ellipticity measurements, leading to higher uncertainty in shear estimation. Additionally, algorithms used for shear measurement can be significantly affected by very noisy sources, introducing noise bias. To address this, galaxies are typically filtered based on their magnitude or S/N. In addition to straightforward selection criteria, weights are often assigned to individual galaxies based on the variance of shape noise and ellipticity measurement noise (e.g., \citealt{miller2013model,jarvis2016,fenech2017selection}). Noisy measurements with high variance are generally given lower weights to account for their impact on the analysis.

\cite{tewes2019ml} proposed the incorporation of a successive network after calibration to predict the weights of different galaxies based on their measured features. We adopted the same ML approach for weight assignment. Similar to the training process for calibration NNs, the NNs predicting weights (hereafter referred to as weight NNs) are trained on sets of galaxies subjected to different constant shears. Each galaxy's weight is determined based on the galaxy's features as measured by the CNN, the same inputs in the calibration process. The relation between the predicted weights and the galaxy features is learned by the NN by minimizing the loss function, which is computed as the squared difference between the weighted summation of shear estimates over the true shear, given by
\begin{equation}
    {\rm MSWB}=\frac{1}{n_{\rm case}} \sum_{i=1}^{n_{\rm case}}[\frac{\sum_{j=1}^{n_{\rm rea}} g_{ij}(\textbf{\textit{f}})w(\textbf{\textit{f}})}{\sum_{j=1}^{n_{\rm rea}} w(\textbf{\textit{f}})} - g_i^{\rm true}]^2
.\end{equation}
Here, $g_{ij}(\textbf{\textit{f}})$ represents the calibrated shear point estimate, and $w(\textbf{\textit{f}})$ denotes the weight prediction, confined to the (0,1) range by the activation function of the output layer. The weight NNs are trained subsequently to train the calibration NNs. We employed eight independent NNs, each consisting of one hidden layer with five nodes, to minimize the aforementioned loss function. The final weight is determined by averaging the values from the four best-performing networks. To carry out both shear calibration and weight prediction, we utilized the publicly available code \textit{tenbilac}\footnote{\url{http://cdsarc.u-strasbg.fr/viz-bin/qcat?J/A+A/621/A36}} developed by \cite{tewes2019ml}. Details regarding the training and validation datasets are provided in Sec. \ref{data}.

\section{Datasets}
\label{datasets}

In this section, we provide a comprehensive overview of the simulations employed in our research, which encompass galaxies, PSF, and observing conditions specifically tailored for the Chinese Space Station Telescope. Additionally, we outline the structure and arrangement of our training datasets for various components of our algorithm.

\subsection{CSST imaging simulations}
\label{simulation}

All the data in this work were generated with the CSST simulation code \footnote{\url{https://csst-tb.bao.ac.cn/code/csst_sim/csst-simulation}}, where the imaging part is based on {\it Galsim}. CSST simulation contains an end-to-end pipeline from numerical cosmology simulation and gravitational ray tracing to optical instruments and imaging. 

We simulated our galaxies as pure exponential disks convolved with non-stationary PSFs simulated for CSST (an example is shown in Fig. \ref{fig:psf}). We generated galaxies stamp by stamp with a size of $128\times128$ pixels, with pixel size $0.074~\rm arcsec$. The galaxies are placed with a uniformly random subpixel offset around the stamp center. We assumed a perfectly known PSF on a stamp of $48\times48$ pixels put into the network. The distribution of galaxy parameters (magnitude, galaxy half-light radius, S/N) is shown in Fig. \ref{fig:catalog}. We only considered single-band measurement in this work. 

Instead of parametric models, the PSF was derived based on an optical design model. To generate a set of realistic PSFs to account for the impact of the optical system on image quality, an optical emulator has been developed to simulate high-fidelity PSFs of CSST. The optical emulator of CSST was based on six different modules to simulate the optical aberration due to mirror surface roughness, fabrication errors, CCD assembly errors, gravitational distortions, and thermal distortions. Moreover, two dynamical errors, due to micro-vibrations and image stabilization, were also included in the simulated PSF.

We have included various sources of noise in the simulated images of CSST. This includes shot noise, sky background, and detector effects. To achieve this, we utilized Galsim to generate photons from a given galaxy, taking into account the throughputs of the CSST system. These throughputs encompass the mirror efficiency, filter transmission, and quantum efficiency of the detector. Additionally, we introduced Poisson noise originating from the sky background and the dark current of a CCD detector. The i-band background level was set to 0.212 e$^-$/pixel/s, while the dark current amounted to 0.02 e$^-$/pixel/s. This results in an average of approximately 35 $\rm e^-$/pixel in a 150s exposure. Furthermore, we incorporated read noise by applying a Gaussian distribution with a standard deviation of around 5.0 e$^-$/pixel. To simulate the production of mock galaxy images on the detector, we also considered bias and applied a gain factor.

\subsection{Data organization}
\label{data}

\begin{figure}
    \centering
    \includegraphics[width=\linewidth]{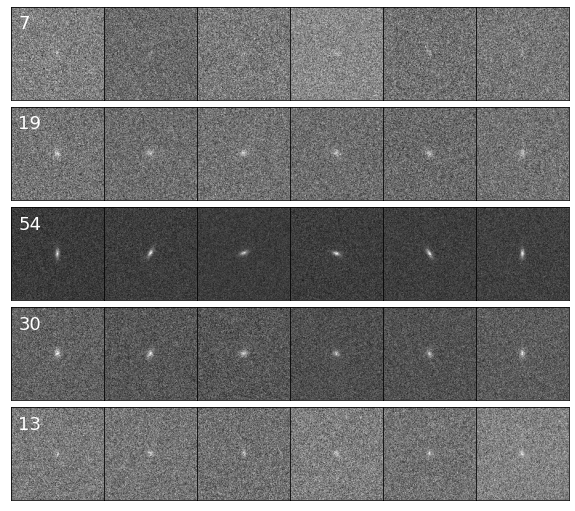}
    \caption{Structured data set to train the calibration NN with the MSB loss function (similar to Fig.~2 in \citealt{tewes2019ml}). Each row corresponds to one case containing 2000 galaxies (i.e., 2000 columns), which differ only in the orientations sharing the same shear and PSF. 5000 cases (i.e., 5000 rows) in total are used to train the NN. Shape noise cancellation is adopted.}
    \label{fig:nn_data}
\end{figure}

\begin{figure*}
    \centering
    \includegraphics[width=0.96\textwidth]{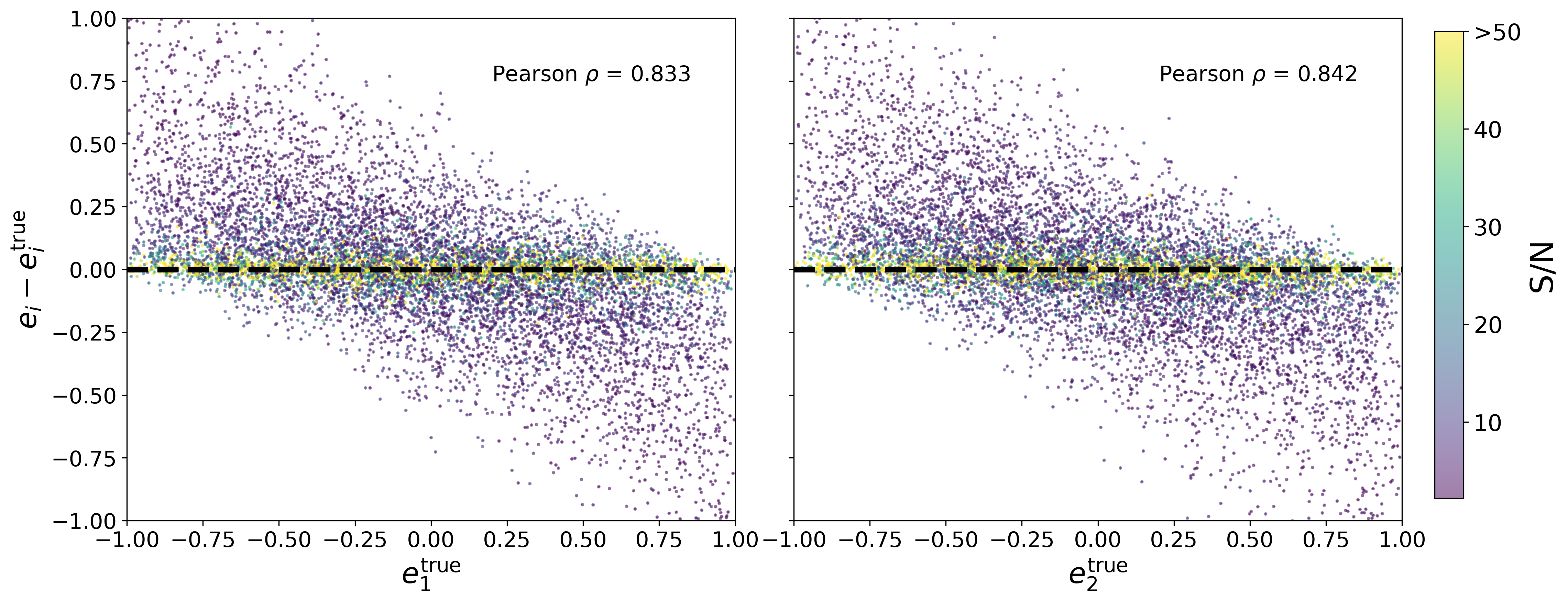}
    \includegraphics[width=0.48\textwidth]{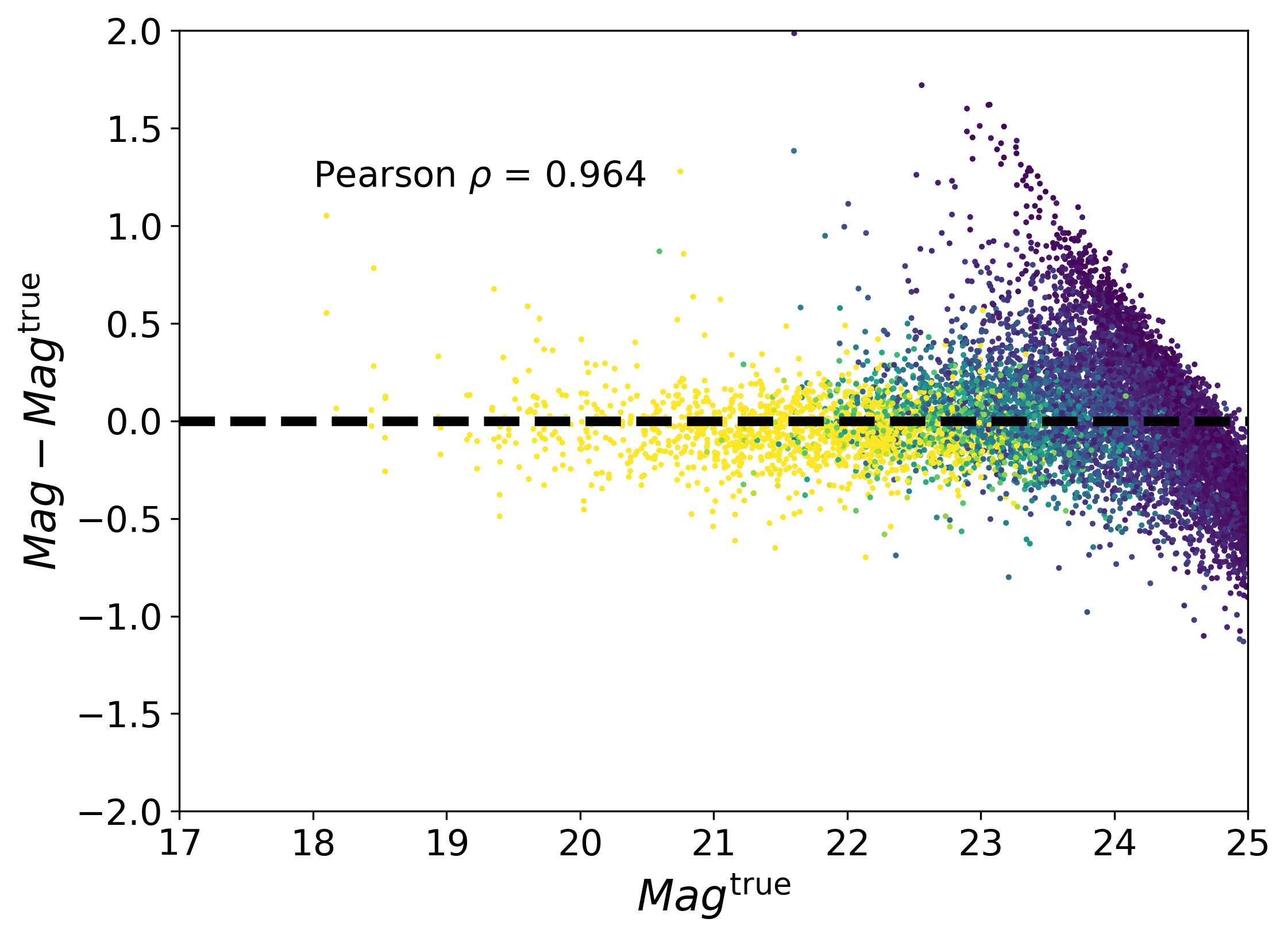}
    \includegraphics[width=0.44\textwidth]{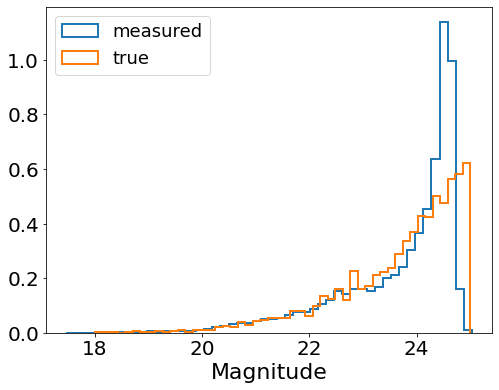}
    \caption{The residuals of CNN measurement on galaxy ellipticities and magnitude. Top Panels: The ellipticities' measurement residuals on 10,000 galaxies of CSST simulations. Colors denote the measured S/N of images, and those with S/N > 50 are shown as the same color as 50. The accuracy sees a strong dependence on the galaxies' S/N. $\rho \simeq 0.98$ for S/N > 10 comparing $\rho \simeq 0.71$ for S/N < 10. Bottom Panels: Measurement residuals of galaxy magnitude in i band (left) and its histogram of truth and predictions (right). Measurements on faint sources ($M_i>24$) are highly biased into being brighter, which propagates into the measured S/N leading to an overestimation.}
    \label{fig:scatter}
\end{figure*}

Galaxy properties including magnitude and half-light radius used in training and analysis were drawn from the CSST catalog, shown in Fig. \ref{fig:catalog}. Throughout the training and evaluation of CNN feature measurements, the input galaxy ellipticities were uniformly drawn within the complex unit circle of $e_1+i e_2$. This deliberate selection ensures that the training data comprehensively cover the parameter space of ellipticity. Our testing has demonstrated that utilizing a nonuniform ellipticity distribution during CNN training results in a notable bias that becomes particularly significant in subsequent shear measurements. For the training and validation of shear measurement, galaxy axis ratios were uniformly sampled from the interval [0.1, 1], while position angles were uniformly drawn from the range of [-$\pi$, $\pi$]. While an alternative plausible approach could involve employing a Gaussian distribution with a dispersion for intrinsic galaxy ellipticities, this difference is not anticipated to have a substantial impact on our ultimate outcomes. Different PSFs at random positions on the CCD were arbitrarily assigned to each galaxy and assumed to be perfectly known when performing measurements. We used a dataset of 200,000 galaxy and PSF pairs in total to train the CNN. 20,000 pairs were used for validation to make sure the model is not overfitting and is generally valid for data not involved in the optimization.  

To train the calibration NNs, we followed the data structure adopted by \cite{tewes2019ml}, in which data were cataloged into "cases" and "realizations" (see Fig. \ref{fig:nn_data}). A realization is a single observation including a galaxy and its PSF. A case is an ensemble of realizations for the same value of a known shear. The training data is grouped into 5000 cases of different magnitudes, galaxy sizes, PSF, and applied shear. 10\% of the cases are separated for validation. Inside each case, there are 2000 realizations sharing the same galaxy and PSF properties (including galaxy axis ratio), and a known shear, but varying in the galaxy orientations. Here, in each case, we adopted the "intrinsic shape cancellation" technique, which ensures the galaxies are perfectly randomly oriented and there are no "shape noise" residuals left averaging their intrinsic ellipticities. More specifically, half of the galaxies were derived by simply rotating 90 degrees of the other, and we then had 1000 pairs of orthogonal galaxies inside each case. Among cases, the shear randomly varies from -0.1 to 0.1, and galaxy properties are sampled from the CSST mock catalog. Different PSFs are also randomly assigned to each case. It is important to mention that galaxies with a half light radius of $r_{50} < 0.1~\rm arcsec$ (which accounts for approximately 5.6\% of the original catalog) are excluded from both the training and validation processes in the subsequent shear measurement. These particular sources exhibit higher shear residuals, which can impact the training phase. While it is theoretically possible to mitigate this effect by incorporating galaxy weighting during the training of weight predictions, we have chosen to exclude these sources from the subsequent tests for the sake of simplicity.

In training the weight NNs, we utilized a dataset consisting of 200 cases, with 10\% of the cases allocated for validation purposes. Unlike the calibration training, each case in this scenario comprises 180,000 pairs of varying galaxies and PSFs sampled from the catalog, all of which share the same shear value within the range of [-0.1, 0.1]. Here, we refrained from applying the shape noise cancellation (SNC) technique. This deliberate choice was made to prevent the NNs from assigning disproportionately higher weights to brighter galaxies with more accurate ellipticity measurements. Doing so would result in a significant loss of information obtained from fainter sources.

 \begin{figure*}
    \centering
    \includegraphics[width=0.45\textwidth]{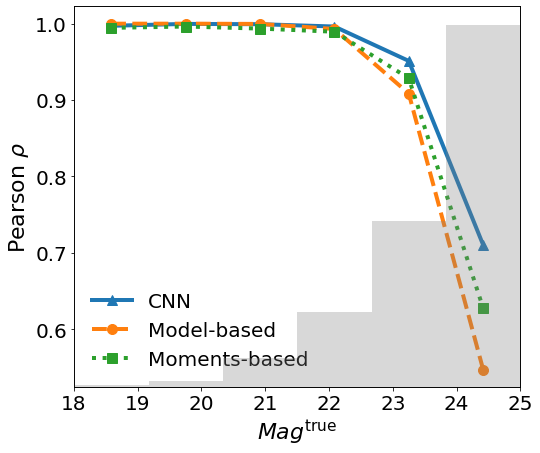}
    \includegraphics[width=0.45\textwidth]{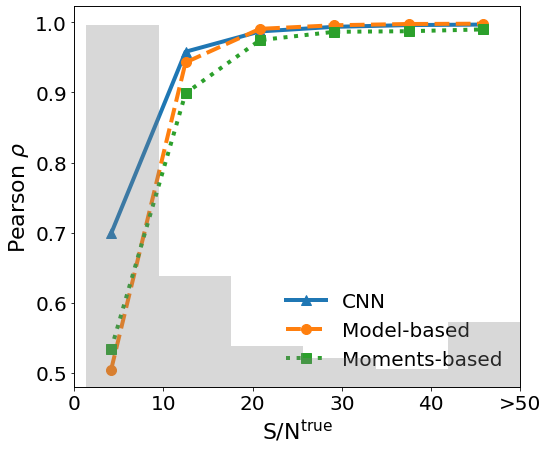}
    \caption{Pearson coefficient of galaxy ellipticity measurements with three methods (CNN; REGAUSS, \citealt{rowe2015galsim}; model fitting, \citealt{sheldon2015ngmix}) as a function of the true magnitude and galaxy S/N. Gray histograms show the galaxy distributions. Galaxies with S/N > 50 sit in the bin of 45-50. The REGAUSS method exhibits failures in feature measurement (on highly noisy or highly elliptical sources), resulting in the rejection of approximately 55\% of the sources. These galaxies are excluded from the $\rho$ calculation in the "moments-based" method, while they are included in the "model-based" and "CNN" methods.}
    \label{fig:compare_1}
\end{figure*}

 \begin{figure*}
    \centering
    \includegraphics[width=\textwidth]{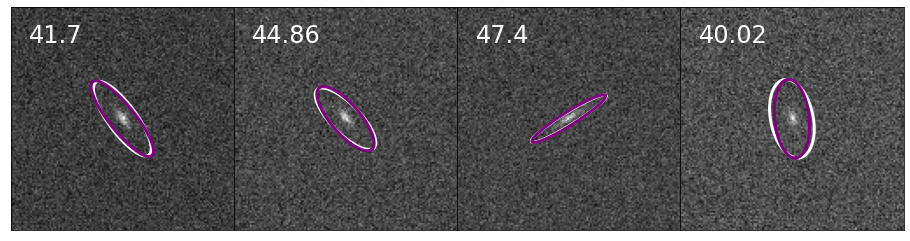}
    \includegraphics[width=\textwidth]{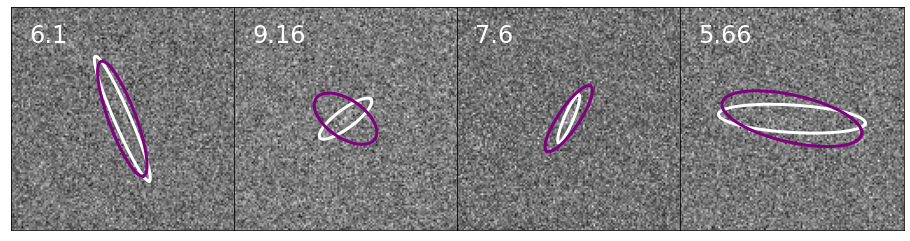}
    \caption{Galaxy shape measurements after PSF correction for eight example galaxies each with measured S/N labeled. The estimates of intrinsic ellipticity and disk half-light radius are shown as ellipses of which the sizes are increased by 12 times for illustration purposes. The white ellipses are ground-true, and the predicted results by CNN are shown in purple.}
    \label{fig:show}
\end{figure*}

\begin{figure}
    \centering
    \subfigure{\includegraphics[width=0.5\textwidth]{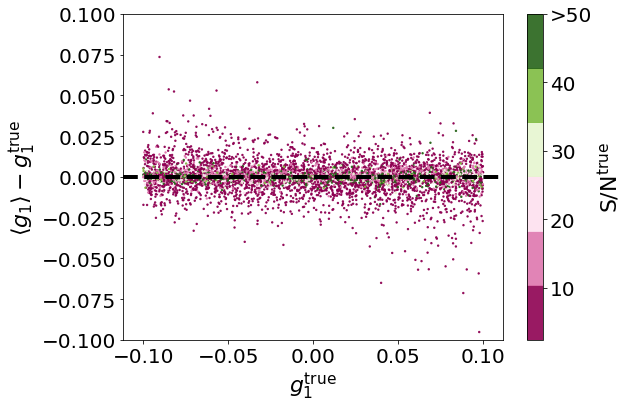}} 
    \subfigure{\includegraphics[width=0.5\textwidth]{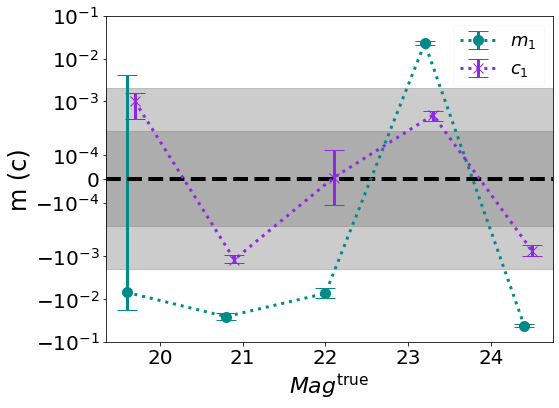}} 
    \subfigure{\includegraphics[width=0.5\textwidth]{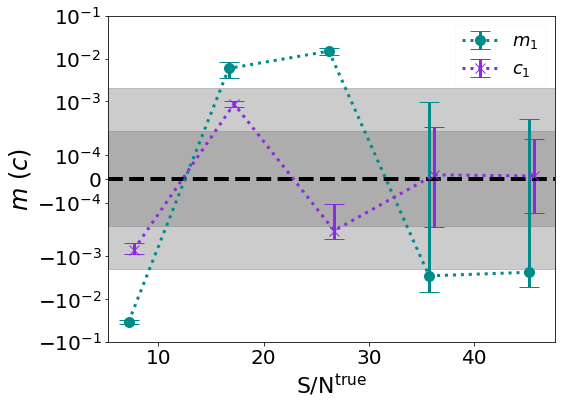}} 
    % \subfigure{\includegraphics[width=0.48\textwidth]{figures/shear_error_mag.png}}
    % \subfigure{\includegraphics[width=0.48\textwidth]{figures/shear_error_size.png}}
    \caption{Shear measurement residuals after calibration and binned shear biases as a function of galaxy properties. Top: Shear estimation on data described in Fig. \ref{fig:nn_data} after NN calibration. Each point is one "case" with 2000 "realizations" sharing the same axis ratio, size, magnitude, PSF, and shear, but differing in orientation. Middle: Multiplicative bias (shown in dark cyan) and additive bias (displayed in blue-violet) are presented as a function of the true galaxy magnitude. The data points from the top panels are categorized into six bins based on the magnitude and fit to a linear function. The y-axis is plotted on a logarithmic scale. The lighter shade corresponds to $\pm 2\times10^{-3}$, while the darker shade represents $\pm 2\times10^{-4}$. Bottom: Similar to the middle panel; the $m$ and $c$ are shown as a function of the true galaxy S/N.}
    \label{fig:shear_train}
\end{figure}

\begin{figure*}
    \centering
    \subfigure{\includegraphics[width=0.48\textwidth]{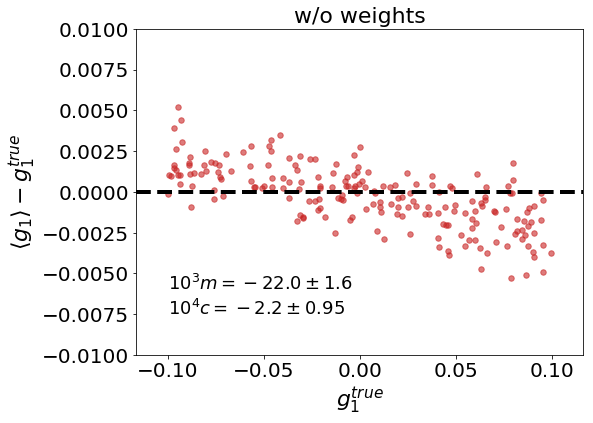}} 
    \subfigure{\includegraphics[width=0.48\textwidth]{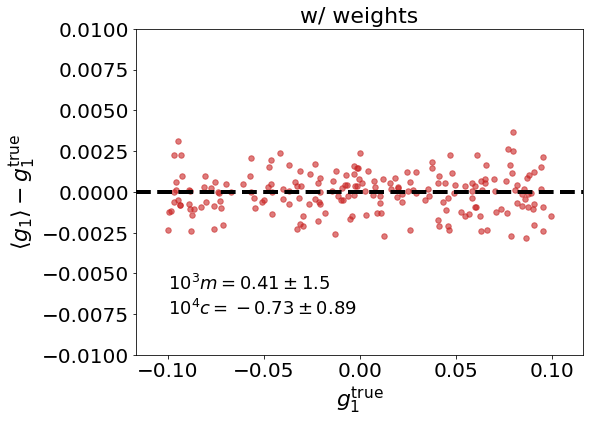}} 
    % \subfigure{\includegraphics[width=0.48\textwidth]{figures/weights.png}} 
    % \subfigure{\includegraphics[width=0.48\textwidth]{figures/weights_snr.png}} 
    \caption{Final results of shear measurement for CSST with our \textsc{Forklens} approach. The first panel displays the shear residuals obtained after calibration using NNs, while the second panel shows the shear residuals after applying galaxy weighting. Both panels consist of measurements from the same set of 20 million galaxies. Each data point represents 100,000 galaxy and PSF pairs with varying properties, but sharing the same shear. Shape noise cancellation was not employed.}
    \label{fig:shear}
\end{figure*}

\begin{figure}
    \centering
    % \subfigure{\includegraphics[width=0.48\textwidth]{figures/shear_g1_cali.png}} 
    % \subfigure{\includegraphics[width=0.48\textwidth]{figures/shear_g1_weight.png}} 
    \subfigure{\includegraphics[width=0.48\textwidth]{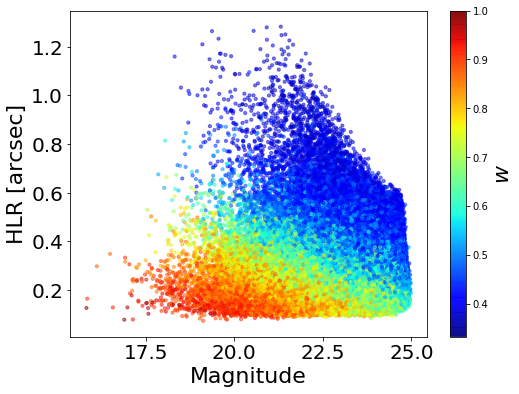}} 
    \subfigure{\includegraphics[width=0.48\textwidth]{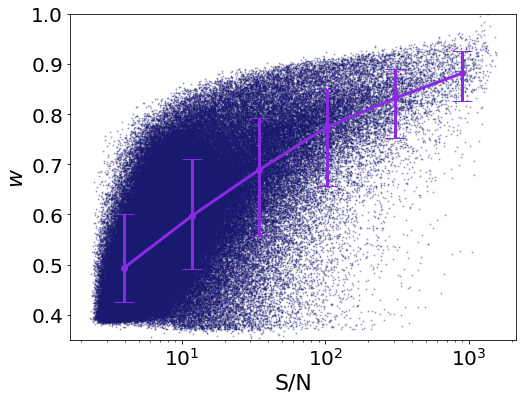}} 
    \caption{Shear weight distributions as function of measured galaxies' properties. In the upper panel, we present the joint distribution of predicted galaxy weights by the NNs, considering the measured galaxy's half-light radius and magnitude as variables. The lower panel illustrates the weights regarding the measured S/N Additionally, the medians of binned weights corresponding to different S/N values are depicted as blue-violet data points, accompanied by their associated 16\% and 84\% errors.}
    \label{fig:weight}
\end{figure}

\section{Results}
\label{results}

In this section, we show the results of the shear estimation with CSST imaging simulation using \textsc{Forklens}. First, we present the results of testing the CNN on CSST simulation, comparing its performance with the moment-based method and model fitting. We then show the shear calibration with NNs based on the outputs of CNN against the results of \textsc{metacalibration}.

\subsection{Feature measurements on CSST simulation}

Our definition of S/N is equivalent to the one adopted by \citet{sheldon2017}:
\begin{equation}
    ({\rm S/N})^2=\frac{\sum I(x,y)^2}{{\rm Var}(I(x,y))}
\label{eq:snr}
.\end{equation}
${\rm Var}(I(x,y))$ is calculated from the edge pixels of a sufficiently large stamp around the galaxy, and $I(x,y)$ is the noise-free elliptical Gaussian model based on the CNN prediction. This definition is similar to the one used in \citet{mandelbaum2014great3}, but with the true profile replaced by the CNN measurement. This S/N estimator is achievable in real data, although it can be a biased indicator compared to other more conventional measurements (e.g., Gaussian aperture or FLUX\_AUTO/FLUXERR\_AUTO from SExtractor outputs \citep{tewes2019ml}). Additionally, the "true" S/N used in this paper refers to the results of using the true galaxy image and noise.

To evaluate the dispersion of shape measurements, we adopted the Pearson correlation coefficient as our metric,
\begin{equation}
    \rho(x,y)=\frac{\sigma_{x,y}}{\sigma_x\sigma_y}
,\end{equation}
where $\sigma_{x,y}$ is the covariance of ground truth and predicted results. $\sigma_x$ and $\sigma_y$ are the respective standard deviations. We provide the estimation of multiplicative and additive bias of shear measurement in Sec. \ref{shear_results}.

The overall performance of the CNN in predicting galaxy features is illustrated in Figure \ref{fig:scatter}. In the top panel of Figure \ref{fig:scatter}, we observe a noticeable correlation between accuracy and galaxy S/N. Bright galaxies exhibit lower dispersion compared to faint galaxies. The Pearson coefficients indicate a correlation of approximately 0.84 (for S/N > 2) in ellipticity measurement and approximately 0.98 (for S/N > 10). Regarding the estimation of galaxy magnitude, the measurements for bright galaxies ($M_i<23$) are accurate, while CNN tends to overestimate the brightness of faint galaxies ($M_i<23$). As depicted in Fig. \ref{fig:catalog}, these offsets in galaxy feature measurements collectively contribute to an overestimation of the S/N when compared to the true S/N calculated using the true, unaltered galaxy instead of the measured Gaussian approximation.

Figure \ref{fig:compare_1} illustrates a comparison of the Pearson coefficient ($\rho$) among three different methods based on galaxy magnitude ($M_i$) and S/N. We divided a sample of 10,000 galaxies into seven groups and calculated $\rho$ within each bin by comparing the predicted ellipticities from each method to the true labels. It is important to note that approximately 52\% of the sample is excluded from the coefficient calculation for the moment-based method due to feature measurement failures with the HSM.shear\_estimate method. However, these galaxies are still included for model fitting and the CNN approach. Consequently, the performance of the moment-based method should be considered comparatively worse in this analysis. The general trend of $\rho$ is similar across the methods, particularly when measuring the brightest galaxies, although the moment-based method performs notably worse. In the case of faint sources with low S/N, the CNN demonstrates better accuracy. In Fig. \ref{fig:show}, we seleted several representative examples showcasing how the CNN recovers the intrinsic shapes of galaxies after PSF correction. The predicted shapes closely match the true input, particularly in the high S/N bin ($40<{\rm S/N}<50$). Even in the low S/N bin ($5<{\rm S/N}<10$), the CNN exhibits reliable recovery of the true shape of the galaxy despite noise contamination. However, some significant shape bias is also evident in these cases, and there are also instances where the input shape is incorrectly recovered.

\subsection{Shear calibration with a neural network}
\label{shear_results}

We input the four features measured by CNN into a NN with two hidden layers of five nodes. Information on the PSF would not be necessary since the CNNs show good results in preprocessing this effect.

We present the overall results of $\left<g_1\right>$ estimation training the NN in Fig. \ref{fig:shear_train}. We do not show the results for $\left<g_2\right>$ as they show similar behavior. We included all S/N ranging down to 2, and over 50\% of the images have an S/N under 10. The NNs demonstrate successful prediction of accurate estimates for the majority of galaxies. However, we do observe a clear trend in measurement residuals for galaxies with an S/N below 10. The middle and bottom panels of the figure present the conditional shear biases as a function of measured magnitudes and sizes of galaxies. The values of m are found to be significant. It is worth noting that no monotonic trends are observed in the data points concerning either magnitude or S/N. The specific behavior of the curves depends on the particular realization of the training datasets. Generally, the medians of the shear biases (m and c) scatter around zero, which becomes significantly smaller when we consider varying galaxies, as shown in Fig. \ref{fig:shear}. However, the presence of a substantial fraction of low-S/N galaxies (S/N < 5) leads to an apparent shear bias (m), necessitating careful selection or weighting strategies to address this issue.

% \textcolor{red}{NN succeeds in predicting accurate estimates for most galaxies, although we notice a clear inclination in measurement residuals for galaxies with S/N < 10.  The middle and right panels depict the 'conditional' shear biases as a function of the measured magnitudes and sizes of galaxies. The values of m are significant. No monotonic trends of the points are observed with respect to either magnitude or half-light radius. The exact tendency of the curves depends on the realization of the datasets. In general, the medians of m and c scatter around zero, which significantly vanish when we consider varying galaxies as will be shown in Fig. \ref{fig:shear}. However, the big fraction of very noisy galaxies (S/N < 5) will put an apparent m, which requires a selection or weighting.}

In Fig. \ref{fig:shear}, we present the main results of this study, which focuses on the overall shear measurement in CSST simulations. The top left panel displays the direct output of the CNN, where each data point represents the averaged $e_1$ measurements of $100,000$ galaxies extracted from Fig. \ref{fig:catalog} under the same shear. A total of 200 points are plotted. Within each point, we did not employ SNC. Even after calibration, the shear measurements still exhibit a significant multiplicative bias of $m_1=-22.0\pm1.6\times10^{-3}$, primarily due to the presence of noisy sources in the dataset. However, upon adopting weights, this bias is considerably improved to $m_1=-0.41\pm1.5\times10^{-3}$ and $m_2=2.3\pm1.6\times10^{-3}$. In both cases, the additive bias $c$ remains accurate at $\sim2 \times 10^{-4}$. In Fig. \ref{fig:weight}, it is evident that the weight values are closely related to the properties of observed galaxies. The resulting weights tend to favor galaxies with higher surface brightness and exhibit a clear dependence on the S/N. The weights are close to one for very high S/N and decrease quickly when S/N is < 10. It is worth noting that the weight distributions depicted in Fig.11 may exhibit distinct tendencies among stages of the training iterations or different training realizations. For example, there are instances where the weights exhibit minimal reliance on the measured galaxy features. While such weights still yield unbiased shear estimations, they lack a meaningful physical interpretation. From a practical standpoint, one could artificially choose a model from different iterations or realizations of training, where the NN-learned weight holds interpretability based on its relation to galaxy features.

\begin{figure*}
    \centering
    \subfigure{\includegraphics[width=0.48\textwidth]{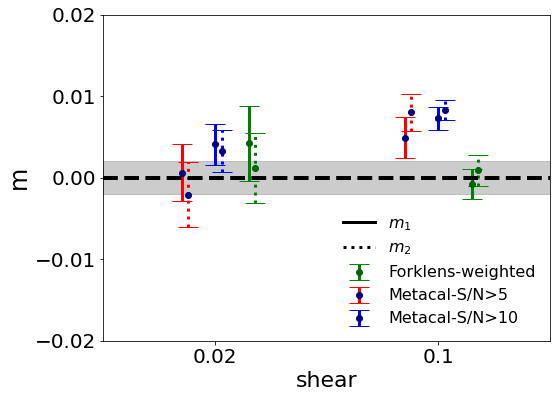}} 
    \subfigure{\includegraphics[width=0.48\textwidth]{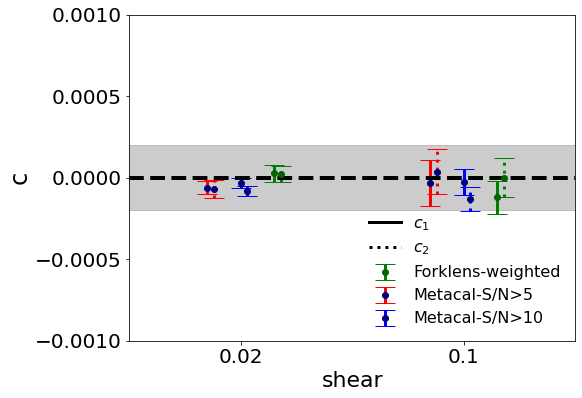}} 
    \caption{Shear measurement bias with \textsc{metacalibration} and \textsc{Forklens} on CSST simulations. Left panel illustrates shear measurement bias $m$, while right panel represents bias $c$ for both \textsc{metacalibration} and \textsc{Forklens} methods on CSST simulation, as a function of the input shear range ([-0.02,0.02] uniformly for a shear of 0.02 and [-0.1,0.1] for a shear of 0.1). Shear biases for the first component are presented with solid error bars, and the second component is depicted with dashed errors. The findings for a shear of 0.02 are obtained from a dataset of 40 million stamps in total (prior to selection) utilizing SNC, and 10 m stamps for 0.1. The gray regions correspond to the requirements set for stage IV weak-lensing experiments, which are $2\times10^{-3}$ and $2\times10^{-4}$ for $m$ and $c$, respectively.}
    \label{fig:metacal}
\end{figure*}

\subsection{\textsc{metacalibration} on the CSST simulation}
\label{METACALIBRATION}

\textsc{metacalibration} operates directly on observed images, with artificial shearing via a series of image manipulations. In Fourier space, the process can be clearly described as
\begin{equation}
    \tilde{I}(\textit{\textbf{g}})=[(\tilde{I}/\tilde{P} \oplus \textit{\textbf{g}})] \times \tilde{P_d}
.\end{equation}
The original galaxy $\tilde{I}$ is deconvolved by its PSF $\tilde{P}$, sheared by an applied shear $\textit{\textbf{g}}$ (in the range of 0.001 to 0.05 and usually 0.01), and reconvolved by a slightly larger PSF than the original one to suppress the amplified noise due to deconvolution. The response of measured ellipticity to shear can then be derived as
\begin{equation}
    R_{i,j} = \frac{e_i^{+}-e_i^{-}}{\Delta g_j}
,\end{equation}
\label{eq:response}
where $e^{+}$ ($e^{-}$) is the measured ellipticity of component $i$ ($i=1,2$) of an image sheared by $+g_j$ ($-g_j$), and $\Delta g_j = 2g_j$. $j$ denotes the two components of shear. The calibrated shear estimation is then a weighted average of
\begin{equation}
    \left<\textit{\textbf{g}}\right> \simeq \left<\textit{\textbf{R}}\right> ^{-1} \left<\textit{\textbf{e}}\right>
,\end{equation}
where $\textit{\textbf{e}}$ is the measured ellipticities on the original galaxy (for more details, please refer to \citealt{sheldon2017}).

This technique has been well tested in the simulation of several surveys and shows a great improvement in shear estimation after calibration (e.g., \citealt{yamamoto2022roman,guinot2022shapepipe}). Not relying on any specific method, \textsc{metacalibration} has the potential and flexibility to be applied to any shape measurement algorithm. Although \textsc{metacalibration} can account for the effect of PSF without any prior PSF correction, it may be beneficial to adopt preprocessing in the case of variable PSFs.

Our CNN can also be directly integrated with \textsc{metacalibration}. \citet{ribli2019shape} combined their CNN with \textsc{metacalibration} and find negligible $m$ and $c$ on DES simulations. Since we used an NN to perform the calibration, we leave this option to future works and stick to a full-ML approach in this paper. However, we do intend to cover how our method compares with \textsc{metacalibration} + model fitting on the same data and see if the strength still stands on noisy images.

% Fig. \ref{fig:metacal} \textcolor{red}{shows} results of \textsc{metacalibration} together with model fitting. We again use the package  \textsc{ngmix} which provides modules for both \textsc{metacalibration} and model fitting. We fit the images with a single Gaussian convolved with a Gaussian PSF. \textcolor{red}{We run this procedure and our CNN+NN (labelled as "Metacal" and "Forklens" respectively in Fig. \ref{fig:metacal}) approach on the same data catalogue. We which are 100 cases of different shears each containing 10,000 galaxies.} Here we employ SNC to better assess the biases introduced by the algorithms in a limited volume of data. We present the measurement bias as a function of the S/N cut in the data, comparing the performance of \textsc{metacalibration} and \textsc{Forklens}. Both "Metacal" and \textsc{Forklens} methods require an S/N selection threshold in our simulation to achieve accurate shear measurements. It is important to note that we use the true S/N to avoid any selection preference for CNN and SNC is applied within each S/N cut.

Figure \ref{fig:metacal} illustrates the comparison between Forklens and \textsc{metacalibration} methods. We again used the  \textsc{ngmix} package, which provides modules for both \textsc{metacalibration} and model fitting. We fit the images with a single Gaussian convolved with a Gaussian PSF. We executed this procedure and our CNN+NN (referred to as Metacal and Forklens, respectively, in Fig. \ref{fig:metacal}) approaches on the same data catalog. Specifically, for shear within the interval of [-0.1, 0.1], each range comprised 100 cases, with each case containing 10,000 galaxies. For shear within the range of [-0.02, 0.02], there were 200 cases, each including 20,000 galaxies. In the latter range, we increased the volume of data to ensure that the errors of m and c were sufficiently small, thus enabling a comprehensive exploration of the shear measurement methods. In the case of Metacal, our simulation necessitated an S/N selection to achieve accurate shear measurements. In comparison, Forklens demanded galaxy weighting, which functioned in a similar manner to a selection process. A selection on the whole galaxy sample (e.g., S/N cut) will modify the distribution of the measured ellipticities, which propagates as a bias into the measured mean shear. The full \textsc{metacalibration} formalism is able to deal with selection effects by calculating a response term similar to Eq. \ref{eq:response}, but accounting for selections \citep{sheldon2017}. For the sake of simplicity, here we employed SNC, avoiding selection bias by counteracting the post-selection shape noise within each shear case. In the selection process for Metacal, we used the true S/N definition.

% The primary feature observed in this comparison would be the larger error bars of $m$ and $c$ for the "Metacal" measurements, which are calculated by fitting the binned shear residuals to a linear function of the true shears. This indicates the lower root mean square (RMS) of \textsc{Forklens} shear measurements, mainly stemming from the superior performance of the CNN in ellipticity measurement on highly noisy images. The error bars decrease with a higher minimum S/N and nearly align at S/N > 15 for both methods, as the CNN strength over model fitting diminishes for bright sources. It is also worth noting that the exact mean positions of $m$ for \textsc{Forklens} vary with different training realizations of the calibration neural networks, which can be attributed to the presence of different trends in the "conditional" $m$ in Fig. \ref{fig:shear_train}. Galaxy weighting can generally account for this issue.

When the shear is at a small magnitude, the response computed in \textsc{metacalibration} can be approximated as linear with respect to the shear. However, this assumption loses validity as the shears grow larger. \textsc{Metacalibration} has been documented to exhibit shear bias that surpasses acceptable limits at higher shears (e.g., $|g| > 0.05$) \citep{sheldon2017, yamamoto2022roman}. As depicted in Fig. \ref{fig:metacal}, the outcomes obtained with \textsc{metacalibration} on CSST simulations align with prior observations. For a shear of 0.1, the Metacal approach introduces a bias of $m_1 = (0.49\pm0.25)\%,m_2 = (0.80\pm0.23)\%$ (S/N > 5), while \textsc{Forklens} demonstrates negligible bias. It reduces to $m_1 = (0.06\pm0.35)\%,m_2 = (-0.21\pm0.40)\%$ for shear of 0.02 using the Metacal technique, in line with $m_1 = (0.42 \pm 0.46)\%,m_2 = (0.12\pm0.43)\%$ observed with \textsc{Forklens}. The c values for both methods remain unbiased in both cases.

Galaxy weighting can be viewed as a simultaneous approach for selection and correction without significant loss in the effective number of galaxies, denoted as $N_{\rm eff}$ \citep{chang2013}:
\begin{equation}
    N_{\rm eff} = \frac{(\sum_i w_i)^2}{\sum_i (w_i)^2}
.\end{equation}
Applying a cut at S/N > 5 (10) preserves approximately 79\% (58\%) of the sources in the entire sample, while the $N_{\rm eff}$ after weighting accounts for approximately 96\% (for $g_1$; 89\% for $g_2$) of the total sources.

In general, our DL-based approach \textsc{Forklens} shows good potential and robustness for stage IV surveys. However, we want to stress that the above comparison is not fair competition. Firstly, we did not include multi-components in the galaxy profile. We did not consider complicated morphology for sources at higher redshifts. Secondly, the data we used in our analysis were perfectly consistent with the simulation we used in training (although they are different). We assumed that our simulations perfectly emulate the real observation, which is not always the case. We leave further comparisons with more complex scenarios for future tests.

\subsection{Time consumption}

As one of the general advantages for nearly all ML methods, CNN measurement is high-speed. With two parallel GPUs and 40 CPU threads (one GPU sees no significant decrease in speed), our CNN takes $\sim 0.7$ milliseconds per galaxy measurement. The time consumption predicting shear point estimates and weights is negligible: $\sim 83$ seconds on 20 million stamps. In comparison, according to our test, \textsc{metacalibration} together with model fitting takes $\sim 0.06$ seconds per galaxy in 40 threads.

Training the models also requires time. It took $\sim 150$ CPU hours with two GPUs to reach the optimized model of CNN. For calibration NNs, the accuracy reaches convergence in $\sim 232$ CPU hours. For weight NNs, it is $\sim 508$ CPU hours. The time required for training the NNs and generating training datasets exceeds that of the actual measurement process. Potentially, this can be significantly accelerated by running on GPUs.

\section{Conclusion and discussion}
\label{conclusion}

We introduced a fully DL-based approach \textsc{Forklens} to measure galaxy shapes and calibrate weak lensing shear. To handle the effect of PSF smearing on observed galaxy shapes efficiently, we developed a two-branch CNN architecture for involving information of galaxy images and PSF simultaneously. Then, we adopted a multilayer NN to calibrate the shear estimate for pixel noise bias. Testing the feasibility of our approach with mock data of CSST, \textsc{Forklens} achieves negligible bias in shear measurement, with $m_1=-0.41\pm1.5\times10^{-3}$ and $c_1=-0.73\pm0.89\times10^{-4}$, based on the analysis of 20 million galaxies. Expectedly, such a setup is suitable for existing and upcoming weak lensing surveys, including both ground- and space-based experiments such as KiDS, DES, {\it Euclid}, {\it Roman}, LSST, and so on.

We employed three different approaches for estimating galaxy shapes: CNN, the moment-based method, and forward-model fitting. The results demonstrate that \textsc{Forklens} exhibits the best overall performance. Specifically, all three methods yield similarly accurate estimations for images with high S/N, but \textsc{Forklens} excels in terms of accuracy when dealing with fainter objects. \textsc{Forklens} also delivers accuracy in shear measurements. When applied to CSST-like mock images, \textsc{Forklens} achieves accuracy on the order of one part in one thousand after incorporating galaxy weighting, which meets the precision requirement of the CSST weak-lensing survey. Compared to \textsc{metacalibration} in conjunction with model fitting, the \textsc{Forklens} approach offers consistent estimations for small shears and yields improved results for larger shears. The weighted effective galaxy number encompasses 95\% of the original sample, thereby preserving a greater amount of information compared to discarding sources with a S/N under five, which would result in retaining 79\% of the sources.

The whole process of our approach is fully automatic, and it costs 0.7 milliseconds for one galaxy if we ignore the time consumption of training, which is much faster than \textsc{metacalibration} integrated with forward-model-fitting shape measurement. However, the time to generate training datasets and train the networks dominates the cost, requiring $\sim$ 890 CPU hours to obtain trained models of CNN and NNs. The automation and efficiency of \textsc{Forklens} make it possible to handle tens of billions of galaxy images from the upcoming large-scale sky surveys with PB-level data.

One highlighted feature of this work is to include separate information in the inputs, which makes it possible for the network to directly learn various effects (in this case, PSF) affecting the target and output corrected results. The concept can be easily extrapolated to other possible situations. One potential example is to use multibranch CNNs to predict galaxy photometric redshift by inputting pixelated images in multiple bands. It might be able to learn all-color morphology and simultaneously predict various properties useful in weak lensing such as shapes and redshifts.
Nevertheless, it is essential to examine both the potential improvements and limitations of our technique. Firstly, it is necessary to acknowledge that we have not addressed the effect of blending. The CNN is trained to operate on individual galaxies centered around the stamp center. The presence of neighboring light contamination due to blending has the potential to impact the current performance of our method. Another critical aspect to investigate is the sensitivity of our method to various factors, including its ability to perform robustly in scenarios that were not accounted for during training. Since we never know the true shear and galaxy shape in real observations, evaluations can only be quantified using simulated images. As a result, the validity of our technique's performance is reliant on the simulations accurately replicating real-world survey conditions.

In the present study, we employed a simplified galaxy profile for all galaxies included in our analysis. However, real-world observations often present a diverse range of galaxy features, such as bulges, knots, and varying light concentrations. Moreover, unforeseen morphologies may also exist, which are not represented in the training dataset. Regarding the PSFs, we assumed that the reconstructed PSF is perfectly known during both the training and testing phases. This means the PSF we feed into the CNN is exactly the same as the one convolved with the galaxy. Nevertheless, even a perfect correction scheme can be subject to systematic biases if the PSF is incorrectly reconstructed \citep{paulin2008psf,jarvis2016leakage}. Additionally, we did not account for various detector effects such as cosmic-ray effects, brighter-fatter effects, and so on, which may present challenges to our current results. Therefore, it is necessary to investigate the impact of these effects on the performance of \textsc{Forklens} in future studies.

Another potential improvement is to implement a Bayesian neural network (BNN) into our model. Common NNs (including the one we use in this work) consist of weights and biases with fixed values. This results in deterministic outputs given fixed inputs. In the Bayesian framework \citep{denker1990bayesian, perreault2017bayesian}, the parameters of the network are instead probability distributions with trainable variances and means. The BNN is able to capture the uncertainties of estimation and provide a confidence level for each output. This might not be necessary for a situation of high consistency between the training and test set. However, considering the above and other unexpected factors not included in the simulation, it is important to know the measurement confidence.

To summarize, we propose a deep-learning-based program (\textsc{Forklens}) to measure weak lensing shears automatically and efficiently for the next generation of large-scale imaging surveys. According to the tests with CSST mock data, \textsc{Forklens} provides better estimation and higher efficiency than traditional methods. By adding more input paths and corresponding PSFs, the \textsc{Forklens} can be easily applicable for other imaging surveys with multi-bands. Currently, we are applying the Forklens to KiDS\footnote{\url{https://kids.strw.leidenuniv.nl/}} and DECam Legacy Survey (DECaLS)\footnote{\url{https://www.legacysurvey.org/decamls/}}. Eventually, we will make Forklens suitable for all fourth-generation imaging surveys.

\begin{acknowledgements}
This work is supported by the National Key R\&D Program of China No. 2022YFF0503403. We also acknowledge the support from the science research grants from the China Manned Space Project with No. CMS-CSST-2021-A01, No.
CMS-CSST-2021-A03, CMS-CSST-2021-A04, No. CMS-CSST-2021-B01 and NSFC of China under grant U1931210.  HYS acknowledges the support from NSFC of China under grant 11973070, Key Research Program of Frontier Sciences, CAS, Grant No. ZDBS-LY-7013 and Program of Shanghai Academic/Technology Research Leader. NL acknowledge the support from CAS Project for Young Scientists in Basic Research (No. YSBR-062). CLW acknowledges the support from NSFC of China under grant 11903082. JY acknowledges the support from NSFC Grant No.12203084. RL acknowledges the support from NSFC Grants (Nos 11988101,12022306), the support from CAS Project for Young Scientists in Basic Research (No. YSBR-062), and the support from K.C.Wong Education Foundation. We use the CSST image simulator to generate the mock data (\url{https://csst-tb.bao.ac.cn/code/csst_sim/csst-simulation}). W.L. acknowledges the support from the GHfund A(202302017475). L.L. acknowledges the support from Natural Science Foundation of Shanghai (No. 21ZR1474200).
\end{acknowledgements}


\bibliographystyle{aa} % style aa.bst
\bibliography{ref} % your references Yourfile.bib

\end{document}